\documentclass[aps,prl,preprint,superscriptaddress,floatfix]{revtex4}
\usepackage{graphicx}
\usepackage{amsmath}
\usepackage{amssymb}

\newcommand{\degree}{$^\circ$}

\begin{document}

\title{Planar array of self-assembled Ga$_x$Fe$_{4-x}$N nanocrystals in GaN: Magnetic anisotropy determined $via$ ferromagnetic resonance}

\author{A.~Grois}
\email{andreas.grois@jku.at}
\affiliation{Institut f\"ur Halbleiter-und-Festk\"orperphysik, Johannes Kepler University, Altenbergerstr. 69, A-4040 Linz, Austria}

\author{T.~Devillers}
\affiliation{Institut f\"ur Halbleiter-und-Festk\"orperphysik, Johannes Kepler University, Altenbergerstr. 69, A-4040 Linz, Austria}

\author{Tian Li}
\affiliation{Institut f\"ur Halbleiter-und-Festk\"orperphysik, Johannes Kepler University, Altenbergerstr. 69, A-4040 Linz, Austria}
\affiliation{Institute of Physics, Polish Academy of Sciences, al. Lotnik\'{o}w 32/46, PL-02668 Warszawa, Poland}

\author{A.~Bonanni}
\email{alberta.bonanni@jku.at}
\affiliation{Institut f\"ur Halbleiter-und-Festk\"orperphysik, Johannes Kepler University, Altenbergerstr. 69, A-4040 Linz, Austria}

\date{\today}

\begin{abstract}
The magnetic anisotropy of a planar array of Ga$_\mathrm{x}$Fe$_{4-\mathrm{x}}$N nanocrystals (NCs) embedded in a GaN host is studied by ferromagnetic resonance.  X-ray diffraction and transmission electron microscopy are employed to determine the phase and distribution of the nanocrystals. The magnetic anisotropy is found to be primarily uniaxial with the hard axis normal to the NCs plane and to have a comparably weak in-plane hexagonal symmetry. The origin of the magnetic anisotropy is discussed taking into consideration the morphology of the nanocrystals, the epitaxial relations, strain effects and magnetic coupling between the NCs.
\end{abstract}

\maketitle

\section{Introduction}

Iron nitride (Fe$_{\mathrm{x}}$N) compounds of various stoichiometries have lately raised much interest as suitable materials for magnetic recording applications \cite{Jack1951,Kim1972,Coey1999,Takahashi2000} and have been extensively studied mostly as polycrystalline thin films and in a powder form.  A particularly interesting iron nitride phase is $\gamma '$-Fe$_{4-x}$N, due to a high spin polarization of conduction electrons \cite{Kokado2006,Tsunoda2010,Tsunoda2009,Ito2012} which makes this material well suited for magnetic write heads and as spin injector in semiconductors \cite{Ando2011,Brataas2002}.  On the other hand, gallium-nitride (GaN) and its compounds are not only strategic semiconductors for opto- and high-power-electronics, but -- when alloyed with magnetic elements -- are emerging as key systems with spintronic functionalities. Moreover, the combination of a nitride semiconducting matrix with embedded magnetic nanocrystals (NCs) is currently opening new frontiers to functional applications. 
Recently, the formation of Ga$_{\mathrm{x}}$Fe$_{\mathrm{4-x}}$N as secondary phases during the metalorganic vapour phase epitaxy (MOVPE) of Fe-doped GaN has been systematically studied and control over the size, density and phase of the Fe-rich species has been obtained by varying the growth conditions and by codoping with acceptors or donors \cite{Bonanni2008,Rovezzi2009,Navarro-Quezada2010,Navarro-Quezada2011}.  A further step was the growth on-demand of planar arrays of Ga$_{\mathrm{x}}$Fe$_{\mathrm{4-x}}$N NCs of a specific crystallographic phase at a defined position in the nitride host \cite{Navarro-Quezada2012}.
 
The determination of the magnetic anisotropy in hybrid semiconductor/magnetic NCs structures is essential for the design of functional devices. In this perspective we report here on the magnetic anisotropy of a planar array of Ga$_x$Fe$_{4-x}$N nanocrystals in GaN at room temperature determined by measuring ferromagnetic resonance (FMR) \cite{Baberschke2011}.  

\section{Experimental Techniques}

The samples studied are grown by MOVPE on $c$-plane sapphire substrates according to a procedure previously reported \cite{Navarro-Quezada2012}.  Specifically, the samples consist of the sapphire substrate, on which a low temperature (530\,\degree~C) nucleation layer is grown, followed by a 1\,$\mathrm{\mu m}$ thick GaN buffer deposited at 980\,\degree C on top of which the Fe-doped layer is grown in a (digital) $\delta$-fashion, $i.e.$ with the Ga flow switched on for 10 seconds and then off for 50 seconds over 30 periods at a temperature of 780\,\degree\,C.  The GaN:$\delta$Fe layer is capped with GaN deposited at 980\,\degree\,C. The precursors employed are ammonia (NH$_3$), trimethylgallium (TMGa) and ferrocene (Fe(C$_5$H$_5$)$_2$), the flow rates during the deposition of the buffer and of the capping layer are 1500 standard cubic centimeters per minute (sccm) for NH$_3$, and 25\,sccm for TMGa.  In the Fe-doped layer the NH$_3$ flow rate is reduced to 800\,sccm, while the ferrocene flow rate is 490\,sccm.  When the TMGa source is enabled for the digital doping, it provides a flow rate of 5\,sccm.

Transmission electron microscopy (TEM) techniques have been employed to establish the crystallographic phase, orientation and distribution of the nanocrystals in the GaN matrix \cite{Navarro-Quezada2012}.  Cross-section and plan-view TEM specimens are prepared by mechanical polishing, dimpling and ion milling in a Gatan Precision Ion Polishing System. Diffraction contrast experiments and high resolution TEM (HRTEM) imaging are carried out in a JEOL 2010F operating at 200\,KeV.

In order to get further insight into the structure and orientation of the nanocrystals, the samples have been also investigated by high resolution x-ray diffraction (HRXRD).  The measurements are performed in a Panalytical X'Pert Pro Material Research Diffractometer equipped with a hybrid monochromator with a 1/4$^{\circ}$ divergence slit. The diffracted beam is measured with a PixCel detector used as single channel detector with an active length of 0.5\,mm, and a 5.0\,mm anti-scatter slit.

The magnetic resonance measurements have been performed with a Bruker Elexsys E580 electron paramagnetic resonance spectrometer capable of static magnetic fields up to 1.5\,T and equipped with a continuous flow cryostat.  An X-band microwave cavity is employed and the measurements are carried out at microwave frequencies between 9.4 and 9.5\,GHz.  The static magnetic field has been modulated with an amplitude of 0.5\,mT at 100\,kHz to allow lock-in detection.  
The samples are cut into 4\,mm$^2$ square specimens necessary to fit the sample space in the spectrometer for in-plane experiments.  The out-of-plane measurements have been carried out by recording an FMR spectrum every 10$^{\circ}$ for one full rotation.  In-plane measurements have been performed every 2$^{\circ}$ for half a circle.  The microwave power has been adjusted to ~2\,mW for the in-plane measurements, while due to larger peak width 20\,mW have been employed for the out-of-plane experiments. 

\section{Experimental Results}

\begin{figure}
  \includegraphics{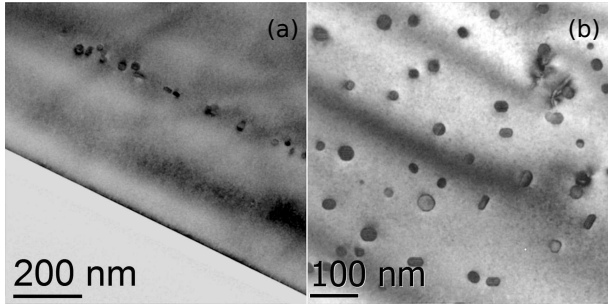}
  \caption{\label{fig:TEM} (Color online) TEM micrographs of the planar array of NCs in the GaN matrix: (a) cross-section, (b) plan-view.}
\end{figure} 

In the TEM micrograph of a cross-section specimen reported in Fig.~\ref{fig:TEM}a, it is shown that the NCs distribute on a plane perpendicular to the $c$ growth direction, over a thickness of $\sim$40\,nm.  Moreover, as estimated from a statistically significant number of plan-view images similar to Fig.~\ref{fig:TEM}b, the average diameter of the NCs is (24.7$\pm$5.2)\,nm.  A similar analysis of 14 cross section TEM images yields an average size of ($16.9\pm2.2$)\,nm along the $c$-direction of GaN and of ($20.1\pm3.9$)\,nm perpendicular to it.  The nanocrystals are estimated to occupy (4.8$\pm$0.2)\% of a basal plane. 

\begin{figure}
  \includegraphics[width=0.5\linewidth]{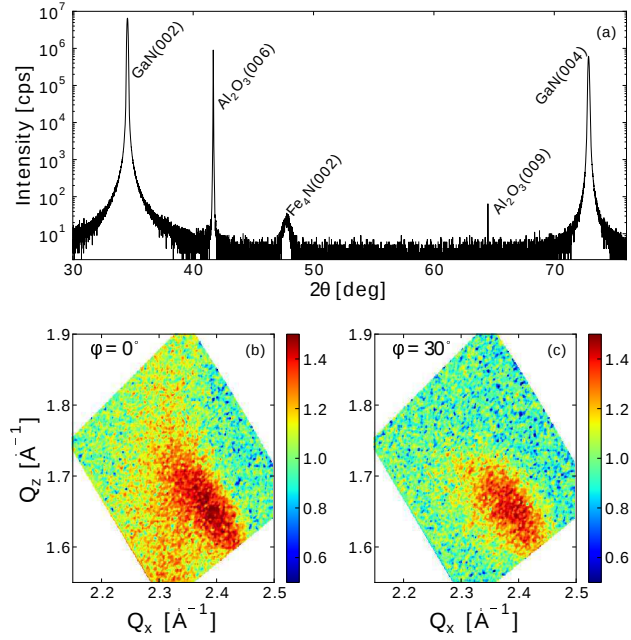}
  \caption{\label{fig:xrd} (Color online) XRD of Ga$_x$Fe$_{4-x}$N nanocrystals in the GaN matrix:  (a) $\theta-2\theta$ spectra, showing the alignement of the (002) of the Ga$_x$Fe$_{4-x}$N nanocrystals with the (002) of GaN;  (b) and (c) reciprocal space maps measured around the (111) peak of Ga$_x$Fe$_{4-x}$N in the (x0z) and (xxz) planes of the GaN reciprocal space ($\phi=0^\circ$ and $\phi=30^\circ$, respectively)}
\end{figure}

The  crystalline phase of the NCs is determined from long radial XRD scans collected along the \emph{c}-axis of GaN as reported in Ref.\,\onlinecite{Navarro-Quezada2012} and established to be $\gamma '$-Ga$_\mathrm{x}$Fe$_{4-\mathrm{x}}$N. Their orientation is determined from the position of the (111) asymmetric peak. The epitaxial relationships between the NCs and the host GaN have been identified as $\left[001\right]_\mathrm{NC}\parallel\left[001\right]_\mathrm{GaN}$ and $\left(001\right)\left[110\right]_\mathrm{NC}\parallel\left(0001\right)\left[11\overline{2}0\right]_\mathrm{GaN}$, giving 12 equivalent in-plane orientations of the NCs, while their basal plane keeps a parallel orientation with respect to the one of GaN.  Moreover, reciprocal space maps of the NCs (111) asymmetric peak have been collected at different azimuths.  A displacement of the $(111)_\mathrm{NC}$ from an azimuth aligned with the $(10\overline{1}0)_\mathrm{GaN}$ to an azimuth rotated by 30$^\circ$ would indicate a uniaxial strain of the NCs, breaking the 12-fold symmetry.  The reciprocal space map of the (111)$_\mathrm{NC}$ collected at four different azimuths taken every 30$^\circ$ and reported in Fig.\,\ref{fig:xrd} does not show any modifications of the peak position in-plane, ruling out a substantial in-plane strain component in the NCs.  Uniaxial strain along the $\left[001\right]$ direction is not clearly detected, yet cannot be excluded throughout.  Furthermore, the XRD measurements indicate that the minority of the nanocrystals is incorporated with the epitaxial relation $\left[111\right]_\mathrm{NC}\parallel\left[0001\right]_\mathrm{GaN}$.  Their number cannot be quantified since their main diffraction peak overlaps with the (006) diffraction of the sapphire substrate.

Temperature dependent FMR measurements have been carried out, and the signal clearly visible at room temperature (RT) broadens with decreasing temperatures, quenching around 40\,K.  Therefore, the anisotropy measurements reported here are restricted to RT.  In this context, effects of thermal broadening may be considered \cite{Charilaou2014} and a similar temperature-dependent behaviour was reported by Bardeleben \textit{et al.} for Co precipitates in ZnO and attributed to the nanocrystalline structure of the films \cite{Bardeleben2008}.

\begin{figure}
  \includegraphics[width=0.5\linewidth]{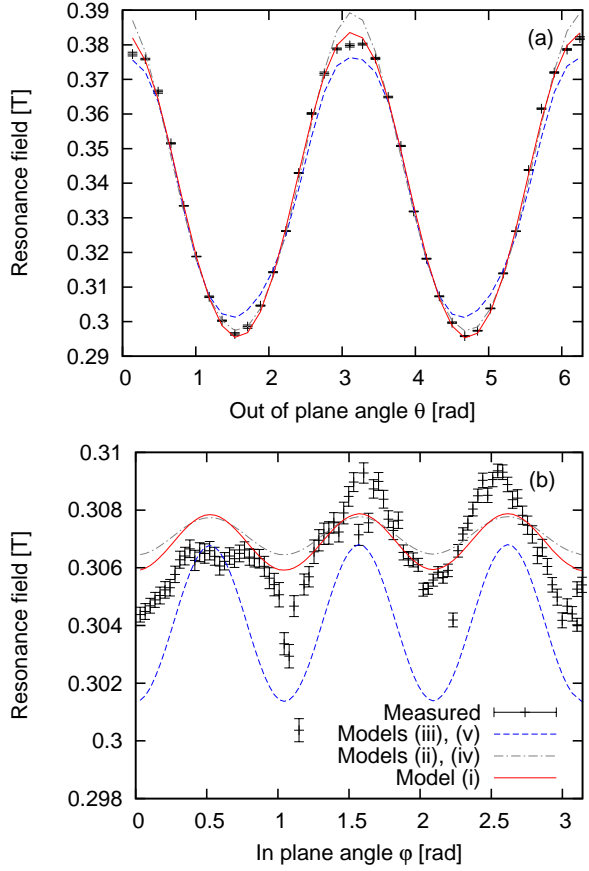}
  \caption{\label{fig:angledependence} (Color online) Angular dependence of the FMR signal (points), plotted together with the fitting results of the models (i)-(v) described in the text. Solid line: phenomenologic crystal field anisotropy; dashed line: fits with cubic nanocrystals with their [111] axis along the [001] axis of GaN, having the uniaxial out-of-plane anisotropy induced by shape anisotropy or coupling, both yielding the same fit quality and lineshape; point-dashed line: fits accounting for the phenomenologic hexagonal in-plane anisotropy and having the out-of-plane uniaxial anisotropy induced by shape or coupling, both giving commensurate fit quality and lineshape.  (a) out-of-plane FMR measurement: the in-plane angle is fixed at $\phi = 0$, and the out-of-plane angle $\theta$ has been varied between 0 and $2 \pi$; (b) in-plane data-points; here the out-of-plane angle is kept at $\theta = \pi/2$ and the in-plane angle $\phi$ is varied between 0 and $\pi$.}
\end{figure}

Due to the equivalent in-plane orientations of the nanocrystals analysed here, one would expect to detect three FMR lines, which coincide when the magnetic field is perpendicular to the sample surface ($\theta=0$).  Nevertheless, only one line can be observed, and shows an uniaxial dependence ($\cos^2\left(\theta\right)$) when varying the out-of-plane angle $\theta$, while measurements with the magnetic field in-plane ($\theta=\pi/2$) evidence a 6-fold symmetry with an angular dependence close to $\sin\left(6 \phi\right)$.    
When the field is nearly perpendicular to the sample surface, the absorption peak is much broader than for angles with the field nearly in-plane.  The angular dependence of the resonance field is reported in Fig.\,\ref{fig:angledependence}.

\section{Discussion}

In order to shed light on the origin of this uniaxial angular dependence, we consider the shape of the nanocrystals as determined by HRTEM, since any non-symmetric uniformly magnetized particle will show uniaxial shape anisotropy.  By approximating the nanocrystals to oblate spheroids \cite{Osborn1945} and using the nanocrystals dimensions mentioned above, the anisotropy tensor in diagonal form has the following components: $N_{xx}=N_{yy}=0.31 \pm 0.03$ and $N_{zz}=0.38 \pm 0.06$.  The error bars of the shape anisotropy do not rule out that in average the NC shape could be isotropic.  In this case, strain induced by the GaN host crystal on the cubic Ga$_\mathrm{x}$Fe$_{4-\mathrm{x}}$N NCs can serve as an explanation for the observed uniaxial anisotropy, since it would generate uniaxial terms in the free energy, which, assuming a high enough prefactor, could dominate over the cubic terms from the iron nitride crystal lattice.  Alternatively, a sizable magnetic coupling between the nanocrystals supported by the planar arrangement would as well lead to uniaxial anisotropy\,\cite{Bardeleben2008}. The coverage required to have coupling energies comparable to the thermal energy at RT can be estimated using the results in Ref.\,\onlinecite{Jensen2003} and correcting the energies reported by the square of the ratio of the saturation magnetization of pure Fe ($1.76\times 10^{6}$\,A/m (Ref.\,\onlinecite{Coey2010})) and Fe$_{4-\mathrm{x}}$N (whose value was experimentally established to be $1.42\times 10^{6}$\,A/m in thin films \cite{Chen1991}, and $1.51\times 10^6$\,A/m in powders \cite{Gillaud1946,Frazer1958,Coey1999}). Based on these figures, the coverage of (4.8$\pm$0.2)\% obtained by (HR)TEM for the samples considered in this work, represents a border limit for the observation of coupling between the NCs at room temperature.

In order to elucidate the origin of the observed in-plane magnetic anisotropy it is mandatory to take into consideration a number of factors.  Shape anisotropy can be ruled out, since for single domain particles only uniaxial like shape anisotropies are enabled.  While the phenomenologic theory of magnetic crystal or strain anisotropy does not exclude the appearance of high order terms, which would not cancel out for the given strain geometry, microscopic arguments speak against strain as possible explanation, since if a quadratic lattice is isotropically stressed along three directions -- with an angle offset of 120$^{\circ}$ between them -- the resulting lattice will still be a square one.  Also, due to the small size of the nanocrystals considered, relaxation of strain is not to be expected, therefore a local change in magnetic anisotropy within a NC is unlikely.
On the other hand the observed 6-fold anisotropy is compatible with the arrangement of a minority of the nanocrystals with their [111] direction aligned parallel to the [001] direction of GaN. Such NCs, together with the uniaxial anisotropy induced by their shape, yield an angular dependence close to the one observed. The angle dependence of the sum of the three peaks due to the other crystallographic orientations nearly averages out (as it can be seen by plotting the lineshape as a function of the angle), while the 6-fold anisotropy of the minority of crystals dominates.
Although, as observed by HRTEM, the interface between the NCs and the GaN matrix is atomically flat, intermixing cannot be excluded. The hexagonal crystal structure of the intermixing region pseudomorphic with wurtzite GaN in the proximity of the NCs would generate an antiferromagntic shell surrounding the NCs and give rise to a hexagonal magnetic crystal anisotropy. For the nanocrystals considered here, the ratio between the atoms at the surface and those in the NC bulk is of the order of 10, thus even a thin antiferromagnetic layer surrounding the nanocrystals is likely to couple to them and to induce a sizable effect. 

The presence of non-interacting paramagnetic Fe$^{3+}$ ions within the GaN host do not show in plane magnetic anisotropy\,\cite{Abragam1970} and can be ruled out as source of the measured effect. On the other hand, the aforementioned possibility of magnetic coupling between nanocrystals even at room temperature would play a role in producing magnetic anisotropy. As stated, high order terms of uniaxial strain anisotropy are possible in the phenomenological description, and if the nanocrystals -- singularly strained along one of the equivalent directions in the host crystal plane -- couple rigidly ($i.e$: with a fixed phase relation of the precessional motion of their magnetization), the overall free energy will contain terms that do not cancel out and that give rise to a hexagonal in-plane anisotropy.  However, the HRXRD data does not provide eveidence of a significant in-plane strain.
In combination with magnetic coupling, also the cubic crystal anisotropy of NCs with their $\left [ 111 \right ]$ direction parallel to the $[001]$ of GaN can describe the observed angular dependence of the FMR signal.

To fit the measured data, five different models have been tested, specifically: (i) fit of a hexagonal crystal anisotropy, without information on the NCs morphology from HRTEM, and yielding apparent anisotropy constants; (ii) an individual treatment of the nanocrystals, including the average dimensions as obtained from HRTEM for the shape anisotropy and adding a phenomenologic hexagonal in-plane anisotropy term; (iii) a model of individual nanocrystals of cubic crystal anisotropy with their $\left [ 111 \right ]$ direction along the $\left [ 001 \right ]$ of the GaN host and taking into account the shape information from HRTEM; (iv) nanocrystals as a rigidly coupled system, $i.e.$ with the shape anisotropy of an infinitely extended thin layer, yet keeping the phenomenologic 6-fold symmetry term; (v) a model treating the whole planar array of nanocrystals as a rigidly coupled system of cubic NCs with the $\left [ 111 \right ]$ direction along the $\left [ 001 \right ]$ of the GaN host.  The formalism employed for the different models is reported in the Additional Material.  In order to determine the resonance frequency, the method described by Smit and Beljers \cite{Baselgia1988} is employed, which allows to calculate the frequency $\omega$ for FMR conditions from the free energy $F$, written as a sum over the different anisotropy contributions and with $\gamma$ as gyromagnetic ratio:

\begin{equation}
\left(\frac{\omega}{\gamma}\right)^2 =\frac{1+\alpha^2}{M^2 \sin \left(\theta\right)^2}\left(\frac{\partial^2 F}{\partial \theta^2}\frac{\partial^2 F}{\partial \phi^2}-\left(\frac{\partial^2 F}{\partial \theta \phi}\right)^2\right)
\end{equation}

In the fit, the equilibrium magnetization direcion ($\theta$, $\phi$) has been determined by minimizing the free energy for each computation of the resonance frequency, while the damping $\alpha$ has been neglected.  All five models describe the observed data well, the best agreement is nevertheless obtained with the purely phenomenologic hexagonal crystal anisotropy model.  The $g$-factor, the saturation magnetization and the anisotropy constants of the crystal anisotropy have been fitted and the values are reported in Table\,\ref{tab:fitresults}.  

Since the uniaxial anisotropy generated by the NC shape, the saturation magnetisation and the uniaxial crystal field anisotropy term are mutually dependent, the obtained quality of the fit is commensurate for the models including the phenomenologic hexagonal term but without the fourth order uniaxial term, and for the two models assuming the cubic anisotropy. The variation is restricted to the fitted anisotropy constant values.

\begin{table*}
  \caption{Material parameters obtained by fitting the angular dependence of the FMR (details are provided in the Additional Material). In the models (i), (ii) and (iii) the nanocrystals are treated as isolated magnetic moments. The model (i) is purely phenomenologic and includes a uniaxial/hexagonal anisotropy, where K$_1$ and K$_2$ are the prefactors of the second and fourth order uniaxial term, and K$_3$ is the prefactor of the sixth order hexagonal term; in (ii) the pre-factors of the uniaxial crystal strain are set to zero, while the model includes the shape information from HRTEM for shape anisotropy; for (iii) the shape information from HRTEM is employed and the nanocrystals are modeled with a cubic anisotropy and with their [111] direction parallel to the [001] direction of GaN, with K$_1$ and K$_2$ being the cubic anisotropy term prefactors of second and fourth order. For the rigidly coupled models the shape anisotropy is set to the one of an infinitely extended thin layer.}
  \label{tab:fitresults}
  \begin{tabular}{p{2.5cm} p{2.5cm} p{2.5cm} p{2.5cm} p{2.5cm} l}
    \hline
    \hline
                           & \multicolumn{3}{c}{Individual} & \multicolumn{2}{c}{Coupled} \tabularnewline
                           & Crystal/Strain & Shape + Hex   & Shape + Cube   & Hexagonal    & Cubic  \tabularnewline
													 & (i) & (ii)   & (iii)   & (iv)    & (v)  \tabularnewline
    \hline
    $g$                    & 2.069          & 2.069         & 2.065         & 2.069        & 2.065 \tabularnewline
    m$_\mathrm{sat}$ [kA/m] & 1443$\pm$12    & 695$\pm$5     & 589$\pm$8     & 49$\pm$0.4   & 41.7$\pm$0.6 \tabularnewline
    K$_1$ [J/m$^3$]        & -40492$\pm$495 & set 0         & 729$\pm$170    & set 0        & 51$\pm$12 \tabularnewline
    K$_2$ [J/m$^3$]        & -3408$\pm$251  & set 0         & -7768$\pm$1000 & set 0        & -551$\pm$71 \tabularnewline
    K$_3$ [J/m$^3$]        & -74$\pm$11     & -24$\pm$8     &                & -1.7$\pm$0.6 &   \tabularnewline
    Fit Error              & 57.46          & 91            & 165            & 91        & 165
    \tabularnewline
    \hline
    \hline
  \end{tabular}
\end{table*}

While the obtained $g$-factor differs from the value reported for Fe$_{4}$N thin films \cite{Nakamura2003}, in the frame of the hexagonal model which is not including the shape anisotropy the fitted saturation magnetization of $1.44\times 10^6$\,A/m is very close to the literature values $1.42\times 10^6$\,A/m (Ref. \onlinecite{Chen1991}) and $1.51\times 10^6$\,A/m (Ref.\,\onlinecite{Gillaud1946,Frazer1958,Coey1999}). When taking into account the shape anisotropy using the average NC aspect ratio as determined by HRTEM, the saturation magnetization given by the fits is lower than the literature value, yet one has to keep in mind the relatively large error bars for the shape anisotropy tensor components, which allow a broad range of fitted saturation magnetisation values without changing the quality of the fit. In order to obtain the literature value for the saturation magnetization of Fe$_4$N, one would have to set a aspect-ratio of $1.08$, which is within the error bar of the size information from HRTEM.  Also, since the lattice parameters of Fe$_4$N and GaFe$_3$N are comparable, the incorporation of Ga into the NCs, which would lead to a decreased saturation magnetization, cannot be ruled out.  Regarding the models assuming rigidly coupled nanocrystals, the saturation magnetization is much weaker than the one reported for thin films. This may be attributed to the fact that in the assumed arrangement of coupled NCs a substantial fraction of the volume is actually represented by paramagnetic dilute (Ga,Fe)N, as proven by SQUID magnetometry measurements \cite{Sawicki2014}.  By calculating the ratio between the literature value and the measured saturation magnetization and neglecting the paramagnetic contribution, approximately 3\% of the film volume is found to consist of Ga$_\mathrm{x}$Fe$_{4-\mathrm{x}}$N, comparable to the value obtained from (HR)TEM.

As stated above, shape anisotropy can fully explain the out-of-plane anisotropy, yet the average dimensions yield a relatively low saturation magnetization.  Since the model neglecting the shape anisotropy and describing the system with crystal/strain provides a significant agreement between the fitted saturation magnetization and the literature value, an estimation of the strain that would be required to obtain the observed uniaxial anisotropy of $\mathrm{K_1} = (-40492 \pm 495$)\,J/m$^3$ has been carried out. According to Ref.\,\cite{Coey2010}, the relation between strain and anisotropy is given by:

\begin{equation}
  K_1 = \frac{3}{2} \lambda_S E \epsilon
\end{equation}

where $\lambda_S$ is the saturation magnetostriction, $E$ the Young modulus and $\epsilon$ the strain.  

Many of the reported experimental values for the saturation magnetostriction of $\gamma '$-Fe$_{4}$N are questionable, due to a misassignment of the reduction of the measured magnetostriction to $\gamma '$-Fe$_{4}$N, which meanwhile it is attributed to nitrogen contaminated $\alpha$-iron. A reasonable figure, as obtained by densitiy functional theory caculations, is $\lambda_S=-10\times 10^{-6}$ (Ref.\,\onlinecite{Coey1999}). For the elastic modulus several theoretical and experimental values have been reported, and lie in the range btween 159\,GPa and 205\,GPa (Ref.\,\onlinecite{Gressmann2007,Yan2009,Zhao2007,Takahashi2012,Yurkova2006,Zheng1995}).  Based on these figures, the strain necessary to generate the observed uniaxial anisotropy would be between 1.3\% and 1.7\%.  Such values cannot be excluded by our HRXRD data. Nevertheless, the most realistic picture is the one considering the presence of a minority of NCs with their [111] direction along the [001] direction of the GaN host.

\section{Summary and Outlook}

In summary, measurements of room temperature FMR of a planar array of $\gamma '$-Ga$_\mathrm{x}$Fe$_{4-\mathrm{x}}$N nanocrystals in a GaN host reveal a strong uniaxial magnetic anisotropy with three easy axis in the plane normal to the \textit{c}-axis of GaN.  The uniaxial anisotropy can be described by the shape of the nanocrystals, and the hexagonal in-plane anisotropy can be understood by considering the presence of a minority of cubic nanocrystals with their [111] axis aligned along the [001] direction of the GaN host, whose presence is indicated by XRD. The data can also be described within a conventional hexagonal magnetic crystal anisotropy model, and a significant agreement is obtained for a layer of coupled crystals.

A broad range of applications for metallic and magnetic nanocrystals in a semiconductor matrix is envisaged and among the most thrilling prospects one can mention the option of exploiting them as spin current injectors into the semiconductor host crystal \cite{Ito2012}, possibly $via$ FMR spin pumping \cite{Ando2011,Brataas2002}, and spin current detection by inverse spin Hall effect \cite{Chen2013}.  A further application is directed to electric flash memory-like data storage \cite{Jeff2011,Li2008}. For magnetic storage in-plane anisotropy is required, and it can be expected to be induced by $e.g.$ elongated nanocrystals obtained by modulating the growth conditions \cite{Devillers2014}. Moreover, the control of the magnetic coupling between nanocrystals makes the hybrid semiconductor/NCs system an exciting candidate to study frustrated magnetic systems and spin glass behaviours \cite{Jensen2003}.

\vspace{\baselineskip}
The work was supported by the European Research Council (ERC Advanced Grant 227690), by the Austrian Science Fundation (FWF Projects 22477, 24471) and by WRC EIT+ within the project NanoMat (P2IG.01.01.02-02-002/08) co-financed by the European Operational Programme Innovative Economy (1.1.2).

\bibliographystyle{apsrev}
\bibliography{./bibliography_nourls_120614}

\end{document}


\section{Additional Material}
    The FMR frequency is derived from the models using the formula of Smit and Beljers \cite{Baselgia1988}, evaluated at the equilibrium magnetization direction.
    The free energy used in the models (i)-(v) fitted to the data reads as follows:
    \begin{equation}
      F=F_\mathrm{Z} + F_\mathrm{S} + F_\mathrm{C}
    \end{equation}
    where $F_\mathrm{Z}$ denotes the Zeeman energy, $F_\mathrm{S}$ the shape anisotropy energy, and $F_\mathrm{C}$ the crystalline or strain anisotropy.  Additional terms, like surface- and step-anisotropy are not taken into account.
    
    The three contributions are given by:
    \begin{equation}
      \label{eqn:fzeeman}
      F_{\mathrm{Z}} = -\mu_0 \vec M \cdot \vec H
    \end{equation}
    Here $\mu_0$ is the Bohr Magneton, $\vec M$ is the magnetization and $\vec H$ is the applied external field.
    \begin{equation}
      F_{\mathrm{S}} = \frac{\mu_0}{2} \vec{M}^\top \cdot \underline{\underline{N}} \cdot \vec M
    \end{equation}
    The tensor $\underline{\underline{N}}$ is the shape anisotropy tensor, rotated to match the coordinate system of the crystal anisotropy.
    
    The formula to be employed for crystal anisotropy depends on the symmetry of the crystal lattice.  Here, cubic and hexagonal anisotropies are considered.  These anisotropies are generally expressed in a phenomenological model which is built using a Taylor expansion of the free energy with respect to the direction cosines $a_i$.  For cubic and hexagonal crystals the first few non-vanishing terms are:
    \begin{equation}
      F_{\mathrm{C}_\mathrm{Cubic}} = K_1 \left(a_1^2 a_2^2 + a_1^2 a_3^2+ a_2^2 a_3^2\right) + K_2 \left(a_1^2 a_2^2 a_3^2\right)+ K_3 \left(a_1^2 a_2^2+ a_1^2 a_3^2+ a_2^2 a_3^2\right)^2
    \end{equation}
    \begin{equation}
      F_{\mathrm{C}_\mathrm{Hex}} = K_1 \left(a_1^2+a_2^2\right) + K_2 \left(a_1^2+a_2^2\right)^2 + K_3 \left(32 \frac{a_1^6}{a_1^2+a_2^2}-48 a_1^4 + 18 a_1^2 \left(a_1^2+a_2^2\right) - \left(a_1^2+a_2^2\right)^2\right)
    \end{equation}
    The labelling of the development coefficients $K_i$ can be different in literature.
    While the above formulation is well suited for the cubic anisotropy, for hexagonal anisotropy it is more illustrative to write it as a function of the angles $\theta$ and $\phi$, defined by the relation between cartesian and spherical coordinates.
    \begin{equation}
      F_{\mathrm{C}_\mathrm{Hex}} = K_1 \sin\left(\theta\right)^2 + K_2 \sin\left(\theta\right)^4 + K_3 \sin\left(\theta\right)^4 \cos\left( 6 \phi \right)
    \end{equation}

    Strain anisotropy is formally equivalent to uniaxial shape anisotropy (first term of the hexagonal anisotropy) and will not be covered separately, considering that the link to strain has already be mentioned in the main text.  

    The models used in the text are:
    \begin{itemize}
      \item (i) hexagonal crystal anisotropy:
        \begin{equation}
          F=F_\mathrm{Z} + F_\mathrm{C_{Hex}}
        \end{equation}
      \item (ii) hexagonal crystal anisotropy + Shape information from HRTEM:  The shape anisotropy tensor is calculated from the size information obtained by HRTEM, using the formulas for an oblate spheroid \cite{Osborn1945}.
        \begin{equation}
          F=F_\mathrm{Z} + F_\mathrm{S} + K_3 \sin\left(\theta\right)^4 \cos\left( 6 \phi \right)
        \end{equation}
      \item (iii) cubic crystals with their [111] direction along [001] of the GaN host:
        the same shape anisotropy tensor as above, rotated with respect to the crystal axis with Euler angles $-\pi/4$, and $-\arcsin\left(\sqrt{2/3}\right)$, while the third Euler angle is the in-plane rotation of the crystals, which is adjusted to get the best agreement between measurement and model.
                \begin{equation}
          F=F_\mathrm{Z} + F_\mathrm{S} + F_\mathrm{C_{Cubic}}
        \end{equation}
      \item For the rigidly coupled models (iv) and (v) the shape anisotropy is replaced by the shape anisotropy tensor for an infinitly extended thin film, having only one component $N_{zz}$.  The formalas from above are employed.
    \end{itemize}

\bibliographystyle{apsrev}
\bibliography{./bibliography_nourls}